\documentclass{eas}
\usepackage{graphicx}
\usepackage{epstopdf}
\usepackage[normalem]{ulem}
\usepackage{natbib}
\usepackage{color}

\begin{document}

\title{Mean flow evolution of saturated forced Shear Flows in Polytropic Atmospheres} 
\runningtitle{Mean flow evolution of forced shear flows}
\author{V. Witzke}\address{Department of Mathematics, City University of London, Northampton Square, London, EC1V 0HB, UK}
\author{L. J. Silvers}\sameaddress{1}
%
%
\begin{abstract}
In stellar interiors shear flows play an important role in many physical processes. So far helioseismology provides only large-scale measurements, and so the small-scale dynamics remains insufficiently understood. To draw a connection between observations and three-dimensional DNS of shear driven turbulence, we investigate horizontally averaged profiles of the numerically obtained mean state. We focus here on just one of the possible methods that can maintain a shear flow, namely the average relaxation method.\ We show that although some systems saturate by restoring linear marginal stability this is not a general trend.\ Finally, we discuss the reason that the results are more complex than expected.  
\end{abstract}
\maketitle

\section{Introduction}
The complex gas dynamics present in stellar interiors is important for many stellar process, such as mixing behaviour (see e.g. \citealp{Zahn_1974, 1977A&A....56..211S}) and magnetic field generation \cite*[e.g.][]{annurev.010908.165215, 5188151520100501}. The tachocline, which is located at the base of the convection zone, is believed to play a crucial role in these processes. This thin region  with a strong radial shear flow was theoretically predicted \citep{1992AandA265106S} and subsequently confirmed by helioseismic observations \cite[e.g.][]{1996ApJ...469L..61K}. \\

Velocity measurements obtained by helioseismology suggest a hydrodynamically stable tachocline \citep{CBO9780511536243A016} i.e.\ the approximated Richardson number is significantly greater than the theoretical stability threshold of 1/4 \citep{Miles1961}. However, helioseismology is restricted  to large-scale time-averaged measurements \citep{CBO9780511536243A013}, such that turbulent motions can still be present on small length- and time-scales. Such a scenario, where a hydrodynamically unstable system appears stable on large scales was, for example, suggested by \citet{0004-637X-690-1-783} and remains to be investigated. \\

Due to the inability of observing most astrophysical shear regions in detail, analytical and numerical techniques have to be used to investigate the motions present in such regions. Most local numerical studies of shear flows that lead to turbulence exploit an unforced flow \citep[e.g.][]{2000JFM...413....1C,  2003JPO....33..694S}, which results in a finite lifetime of an initially unstable background state. However, astrophysical shear flows can be either transient features or be sustained over very long time-scales. Thus investigations of astrophysical shear flows use different methods to reach a sustained flow. While we previously have compared various methods to sustain large-scale shear flows during the saturated phase \citep[see][]{Witzke03082016} here we concentrate on just one method. The method selected, the relaxation method, is suitable for modelling a target flow in the saturated phase. This method allows the time-scale on which the system is driven to the initial shear flow to be adjusted. Through our investigations we shed light on the question of how likely it is that an initially unstable shear flow will result in global flow profiles that suggest a stable system. We analyse the saturated regime of two differently stratified systems in terms of their horizontally averaged profiles and the resulting effective Richardson number.  

\section{Model}
\label{sec:Model}
We consider a three-dimensional domain of depth $d$, bounded by two horizontal planes located at $z=0$ and $z= 1$, and periodic in both horizontal directions. The fluid is assumed to be an ideal monatomic gas with the adiabatic index $\gamma= c_p/c_v = 5/3$ and constant transport coefficients. The set of dimensionless equations describing the problem and the forcing method used to sustain an initial velocity, $\mathbf{u_0} = (u_0(z),0,0)^T$, are exactly the same as described in \citet{Witzke03082016}.\\

In order to investigate whether an initially unstable system can reach a saturated state where the horizontally averaged profiles, associated with large-scale averaged measurements, suggest a stable system we focus on two differently stratified cases. Case I is strongly stratified, but the polytropic index is chosen such that it is not far from being unstable to convection. This case was investigated in \citet{Witzke03082016}, where different forcings were compared. Here rather than focusing on the different forcing methods we will instead examine the horizontally averaged profiles during the saturated regime in order to understand if it suggests a stable or unstable system. Case II is weakly stratified, but has a large polytropic index to ensure that the system is far from the onset of convection. The Prandtl number is taken $\sigma = 0.1$ for both cases, as $\sigma < 1$ is more relevant for stellar interiors.  All relevant parameters are summarised in Table \ref{table:parameters}.
Using the relaxation method we consider different relaxation times, $\tau_0 $, in order to investigate how horizontally averaged profiles are affected. Then, the system is evolved sufficiently long after saturation to reach a statistically steady state.
\begin{table}[h]
\centering
\caption{Parameters for the investigated cases, where the resolution of the domain is given by $N_x$, $N_y$ and $N_z$. The dynamical viscosity is $C_k \sigma$, where $C_k$ is the thermal diffusivity and $\sigma$ the Prandtl number. The temperature gradient is denoted by $\theta$ and the polytropic index is m. For the initial velocity profile the shear amplitude is $U_0$ and the shear width is controlled by $L_u$. The resulting minimum Richardson number, $Ri$, is calculated for the initial state.}
\begin{tabular}{l c c c c c c c c c}

\hline
\textbf{Case} & $C_k  \sigma$ & $\theta$ & m  & $U_0$ & $1/L_u$ & $N_x$ & $N_y$ & $N_z$ & $Ri$\\[3pt]
\hline
Case I & $10^{-4}$ & 5 & 1.6 & 0.2 & 80 & 256 & 256 & 360 & 0.003\\
Case II &  $10^{-5}$ & 0.25 & 4 & 0.05 & 40 & 256 & 64 & 384 & 0.07\\

\end{tabular}
\label{table:parameters}
\end{table}
%
%
\section{Results}
\label{sec:effective_Ri_number}
\begin{figure}
\centering
\includegraphics[width=0.49\textwidth]{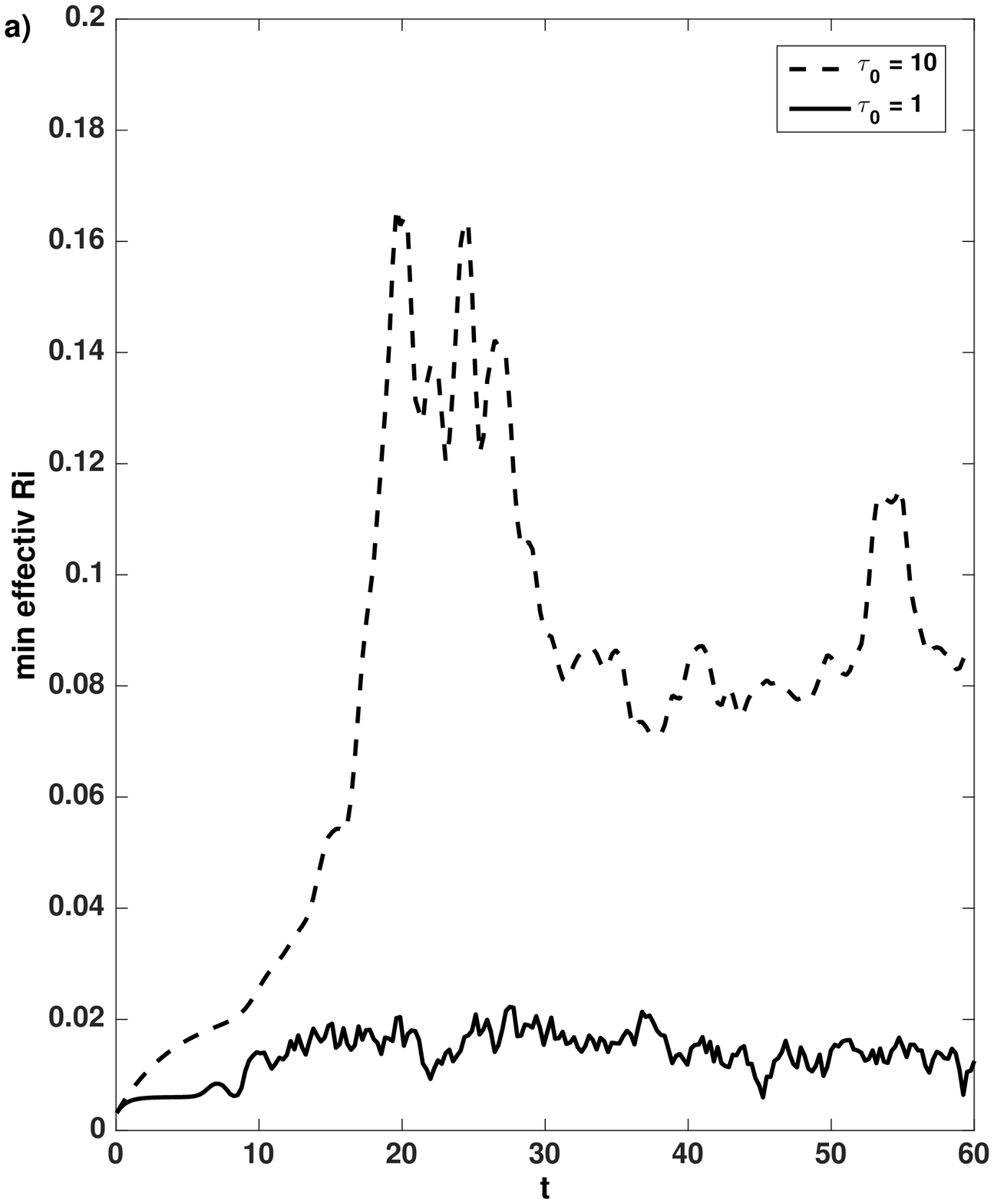}
\includegraphics[width=0.49\textwidth]{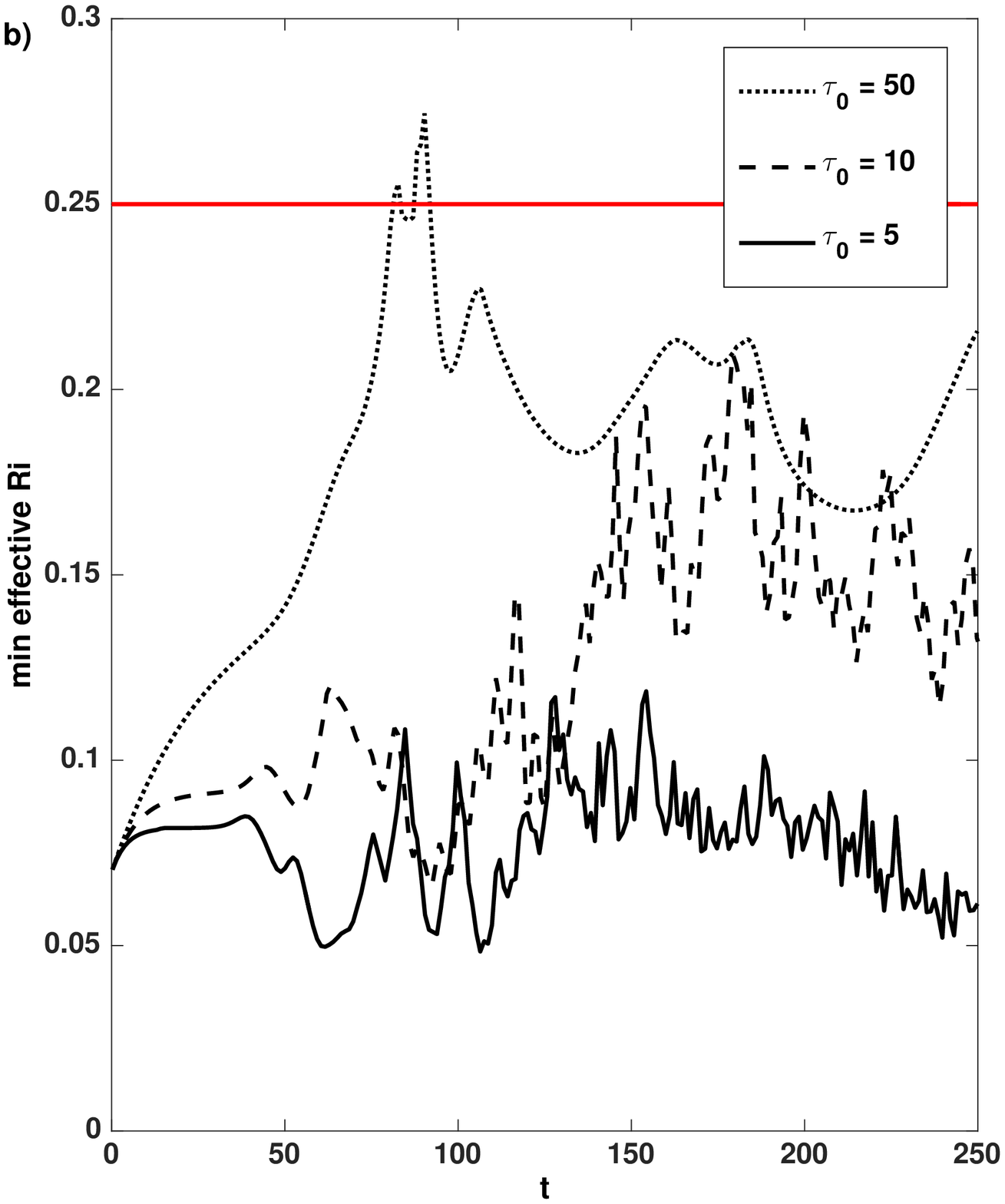}
\caption{The minimal effective Richardson number obtained from the horizontally averaged profiles as in Equation (\ref{eq:effectivRi}) with time for each of the two cases we consider.  a) case I using two different relaxations times $\tau_0$ is displayed.  b) case II with three different relaxation times $\tau_0$ is shown. The red line indicates the 1/4 stability threshold.} 
\label{fig:figure02}
\end{figure}

During the evolution of an unstable shear flow the horizontally averaged density, temperature and velocity profiles are modified. Therefore, the effective minimal $Ri$ number of the system changes. In stratified systems this modification comes from two sources: The change in the Brunt-V{\"a}is{\"a}l{\"a} frequency, due to changes in the averaged density and temperature profiles, and the change in turnover rate
of the shear. For both cases we consider the first contribution remains small compared with the latter one. The minimal effective Richardson number is calculated from the horizontally averaged profiles as follows,
\begin{equation}
\label{eq:effectivRi}
\min Ri_{eff} = \min \left[\frac{-\theta \left(m+1 \right) }{ \left(\frac{\partial \bar{u}(z)}{\partial z}\right)^2} \left( \frac{\gamma -1}{\bar{\rho}(z)}\frac{\partial \bar{\rho}(z)}{\partial z} + \frac{\gamma}{\bar{T}(z)} \frac{\partial \bar{T}(z)}{\partial z} \right)\right],
\end{equation}
where horizontally averaged quantities are denoted by an overbar. This quantity is compared to the initial minimal Ri in order to study the change to the system. 
Investigating $Ri_{eff}$ for case I  (see Fig.\ \ref{fig:figure02} (a)), we find that it increases significantly and reaches a maximum when the instability saturates. Afterwards, $Ri_{eff}$ fluctuates around a value, and this value increases as $\tau_0$ increases. The late time values are $Ri_{eff} \approx 0.09$ for the relaxation method with $\tau_0 = 10$ and $Ri_{eff} \approx 0.01$ for $\tau_0 = 1.0$.  
For both runs of case I, $Ri_{eff}$ obtained in the statistically steady state is one order of magnitude greater than the initial $Ri= 0.003$ number.\\

For case II, see Fig.\ \ref{fig:figure02} (b), the system is initially closer to the stability threshold. Varying the relaxation time gives rise to a similar trend as in case I, where the effective Richardson number during the quasi-static regime increases as $\tau_0$ increases. When using $\tau_0 = 50$, the minimal $Ri_{eff}$ becomes greater than $1/4$ short after the system starts to saturate. However, it drops down to approximately $0.2$ at later times. This shows that for a few special cases the Kelvin-Helmholtz instability saturates by restoring linear marginal stability \citep{1992AA...265..115Z, 2014AA...566A.110P}, but we did not find any evidence for an effective Richardson number greater than the linear stability threshold value during a statistical steady state. 
\section{Conclusions}
Investigating a shear flow instability after the system starts saturating reveals a significant increase in the minimal $Ri$ obtained from horizontally averaged profiles. It becomes evident that the relaxation time used in the forcing method has a significant effect on the resulting minimal $Ri_{eff}$ during the steady state. Starting with an initially unstable configuration close to the stability threshold can lead to a transient phase where the effective $Ri$ number becomes greater than 1/4. However, the minimal $Ri_{eff}$ decreases notably below the stability threshold at later times. It is therefore difficult to achieve a turbulent flow that looks stable in a pure hydrodynamical system with a connectively stable stratification when starting from an initially unstable configuration.\\

\section*{Acknowledgements}
This research has received funding from STFC and from the School of Mathematics, Computer Science and Engineering at City University London. This work used the ARCHER UK National Supercomputing Service (http://www.archer.ac.uk). Some of the calculations were carried out on the UK MHD Consortium computing facilities at Warwick that is supported by STFC.

\bibliographystyle{astron}
\bibliography{bibfile}

\end{document}